# Correlation Statistics for cDNA Microarray Image Analysis

Radhakrishnan Nagarajan*

Center on Aging, University of Arkansas for Medical Sciences
629 Jack Stephens Drive, Room: 3105
Little Rock, Arkansas 72205
Email: nagarajanradhakrish@uams.edu

Meenakshi Upreti

Department of Biochemistry and Molecular Biology

University of Arkansas for Medical Sciences

4301 W. Markham, Little Rock, AR, 72205, USA

Email: upretimeenakshi@uams.edu

*To whom correspondence should be addressed

**Abstract:**

In this report, correlation of the pixels comprising a microarray spot is investigated. Subsequently, correlation statistics namely: Pearson correlation and Spearman rank correlation are used to segment the foreground and background intensity of microarray spots. The performance of correlation-based segmentation is compared to clustering-based (PAM, k-means) and seeded-region growing techniques (SPOT). It is shown that correlation-based segmentation is useful in flagging poorly hybridized spots, thus minimizes false-positives. The present study also raises the intriguing question of whether a change in correlation can be an indicator of differential gene expression.

# 1. Introduction

Microarrays have been widely used to determine the simultaneous expression of genes in distinct biological paradigms [1-3]. They can be broadly classified into single-color and two-color arrays. In the former the control and the experimental specimen are hybridized onto separate arrays, whereas in the latter they are hybridized on to the same arrays [1-3]. The objective is to determine genes that are differentially expressed between two biological states, also referred to as *control* (e.g. normal) and *experimental* (e.g. cancer) samples. These samples are tagged with dyes (Cy3, control) and (Cy5, experimental). The former (Cy3) is green in color with wavelength ~530 nm and the latter (Cy5) is red in color with wavelength ~ 630 nm [3]. The dyes are also referred to as *channels* and can be swapped in the case of a dye-swap experiment [4]. Such an approach can be useful in eliminating spurious differential expression that is an outcome of dye-binding bias as opposed to true biological variability. The tissues tagged with Cy3 with Cy5 are known as *targets* and are hybridized onto the substrate containing the *probes* [5]. Each probe corresponds to a gene. Probes can be either a full length DNA sequences or a short oligonucleotides, and are designed so as to minimize non-specific hybridization. The target is subsequently *hybridized* on to the probes. Hybridization is a complex process and involves several intermediate steps, a detailed description of the hybridization protocol can be found elsewhere [3]. Following hybridization, the arrays are scanned by lasers at two different wavelengths corresponding to the green and red dyes. The dyes bound to the probes fluoresce when scanned at the corresponding wavelength. A detector captures the emitted photons and subsequently converted into an electric current by a photomultiplier tube (PMT). This in turn is digitized into pixel intensities, stored in tagged image file format (.TIFF, 16-bit images). The number of bits representing a grayscale image represents its dynamic range. A 16-bit image has a dynamic range [0, 65535]. A microarray image scanned at wavelengths corresponding to Cy3 and Cy5 is shown in Figure 1. While Spot A represents a gene which is down-regulated in Cy5 with respect to Cy3, spots B and C represent non-differentially expressed genes. Spot B represents a gene that is expressed equally in Cy3 and Cy5 whereas Spot C represents a gene that is expressed neither in Cy3 nor in Cy5, Figure 1. A detailed view of the pixel intensities comprising spots A, B and C is shown in Figure 2.

### 1.1 Segmentation of Microarray Images

Segmentation involves *partitioning* an image into disjoint subsets or regions, such that the pixels within a partition share a common property as opposed to those across partitions [6]. Segmentation of the microarray images is an important preliminary step as any errors incurred at this stage are bound to propagate through subsequent analysis. Consider an image $\mathbf{I}$ partitioned into $k$ regions $R_i, i = 1...k$, then

$$\mathbf{I} = \bigcup_{i=1}^{k} R_i$$

$$R_i \bigcap R_j = f \text{ where } i \neq j \text{ and } i, j \in 1...k$$

The above expressions reflect the *exhaustive* and *exclusive* nature of partitions, i.e. a pixel must be a member of only one of the regions. The choice of the segmentation technique is based on the problem at hand [6]. In microarray image segmentation, the objective is to partition the spot inside a grid into foreground ($\mathbf{F}$) and background ($\mathbf{B}$). Such partitions are also termed as *binary partitions*. Several segmentation techniques have been proposed in the past. These have been broadly classified [7] into (A) Fixed Circle, (B) Adaptive Circle, (C) Adaptive Shape and (D) Histogram method. A concise description of each of these techniques along with their assumptions and references is enclosed in Figure 3. It should be noted that each of the segmentation techniques work under certain implicit assumptions and hence are susceptible to errors when these assumptions are violated.

In the present study, we investigated the *correlation aspects* of pixels comprising a spot using Morgera's covariance complexity [11], subsequently two measures of correlation, namely: Pearson's correlation (parametric) and Spearman rank correlation (non-parametric) [12] are proposed to determine the foreground and background intensity of the given spot. These statistics are also used to flag poorly hybridized spots thus minimizing false-positives. The results of the correlation statistics are compared to three popular microarray segmentation techniques namely: k-means [10], PAM [10] and SPOT [7]. The superiority of SPOT over the other existing segmentation techniques (Section 1.1) is discussed elsewhere [7], hence its choice. The data used in the present study is publicly available [7, 13, 14] and consists of five replicate microarrays (16 bit, .TIFF images) containing the expression of (19 x 21 = 399 genes) generated in a lipid-metabolism experiment [7, 10, 13, 14].

## 2. Correlation statistics for microarray image segmentation

Consider the grid of a microarray spot containing $m$ x $n$ pixels. The number of possible configurations that one can have is: $(m$ x $n)! = (mn)$ x $(mn-1)$ x $(mn-2)$ x … x 2 x 1. In Figure 5, we show a possible configuration ($I^*$) obtained by constrained random shuffling of the rows and columns of a spot $\mathbf{I}$. It is important to note that while the spatial orientation of the pixels of $\mathbf{I}$ is destroyed in $\mathbf{I}^*$, the pixel distribution is preserved. Alternately, segmentation techniques based solely on the distribution of the pixel intensities will be unable to discern $\mathbf{I}$ from $\mathbf{I}^*$. While some of these configurations might resemble a microarray spot, others may not. Thus it might be interesting to investigate the correlation of the pixels comprising a spot. More specifically we address the following questions:

> **QA:** *Do the pixels in Cy3 and Cy5 channels exhibit significant correlation and does this vary with differential expression?*

> **QB:** *Can correlation statistics be used to segment the foreground and background pixels comprising a spot?*

> **QC:** *How does the correlation statistics perform when compared to well-established microarray segmentation algorithms?*

*2.1 Correlation of pixels comprising a spot*

Prior to segmentation of the microarray spot using correlation-based statistics, we address question (**QA**), above. A microarray spot scanned at two different wavelengths corresponding to red (Cy5) and green (Cy3) dyes can be classified under one of the following cases:

> **Case (i)**: *the spot corresponds to a gene that is differentially expressed, i.e. it has abundance of only one of the dyes*.

Up-regulated genes are accompanied by high abundance of Cy5 (red) as opposed to Cy3 (green), whereas down-regulated genes are accompanied by high abundance of Cy3 (green) as opposed to Cy5 (red). Thus for a correlation sensitive statistics, we expect a decrease in its value from highly abundant dye to lowly abundant dye as reflected by a correlation measure.

**Case (ii)**: *the spot corresponds to a gene that is not differentially expressed, i.e. it has equal abundance of both the dyes resulting in yellow color*.

For a correlation sensitive statistics, we expect its value to remain similar between Cy3 and Cy5 channels

**Case (iii)**: *the spot corresponds to a gene that is not differentially expressed, i.e. it has abundance of neither of the dyes*.

For correlation sensitive statistics, we expect its value to remain similar between Cy3 and Cy5 channels.

It is possible to encounter cases where a gene exhibits similar correlation across the Cy3 and Cy5 channels and yet differentially expressed. However, for such spots the intensity of the pixels comprising the foreground (e.g. median foreground intensity, Sec. 2.2) will be significantly different between the channels. Spots (A, B and C) in Figure 2 represent cases (i), (ii) and (iii) respectively. Spot A represents the gene (Apolipoprotein AI) [7, 13, 14] which is down-regulated in the experimental as opposed to the control channel. Scanning Spot A at a wavelength corresponding to Cy3 results in a dense seemingly circular region indicating the abundance of the Cy3, surrounded by a low-intensity region, Figure 2. A gene that is not differentially expressed can have either equal abundance of Cy3 and Cy5 or none. These are represented by spots B and C respectively, Figures 1 and 2. Spot B corresponds to an expressed-sequence tag (EST) [7, 13, 14] expressed equally in the control and the experimental channel, whereas Spot C does not contain any probes (BLANK spot) [7, 13, 14] and is expected to hybridize neither the control nor the experimental channel. Visual inspection of Figure 2 reveals a significant change between the control and the experimental channels for Spot A as opposed to Spots B and C.

Subsequently, we used the Morgera's covariance complexity measure ($h$) [11] was to quantify the correlations of the pixels comprising spots A, B and C, hence address question **QA**. A description of the computational procedure of ($h$) is shown in Figure 4. In the case of two-color experiments, one of the channel acts as the internal control of the other therefore it is important to compare the ratio of the covariance complexities ($h^{Cy5}/h^{Cy3}$) as opposed to its actual value. The ratio ($h^{Cy5}/h^{Cy3}$) was determined for spots A, B and C across each of the five replicate arrays. The choice of replicate arrays is encouraged in microarray studies in order to reject the claim that the observed difference is due to experimental artifacts incurred on a single array. $h^{Cy5}/h^{Cy3}$ estimates for the spots A, B and C estimated across five replicate arrays is shown in Figure 6. It is important to note that ($h$) is inversely proportional to correlation, hence dye abundance. As noted earlier, Spot A represents a gene which is down-regulated, i.e. it is highly expressed in the control (Cy3) as opposed to experimental channel (Cy5). Therefore, ($h^{Cy5}/h^{Cy3} \gg 1$) for Spot A across the five replicate arrays, whereas Spots B and C have a ratio close to one ($h^{Cy5}/h^{Cy3} \sim 1$) indicating similar correlation between the channels. While Spot B exhibits consistently high correlation across Cy3 and Cy5, Spot C exhibits consistently low correlation across Cy3 and Cy5. In order to further justify the existence of correlation, we compared the ratio ($h^{Cy5}/h^{Cy3}$) of spots A, B and C to their random shuffled counterparts. Random shuffles were obtained by a row-wise random shuffle of the pixels followed by a column-wise random shuffle as shown in Figure 5. While the ratio ($h^{Cy5}/h^{Cy3}$) was different between the Spot A, Figure 6, and it's random shuffled counterpart, ratio ($h^{Cy5}/h^{Cy3}$) of spots B and C were similar to that of their shuffled counterpart, Figure 6.

## 2.2 Correlation statistics for spot segmentation:

Having established the fact that the spots comprising a microarray spot exhibit considerable correlation whose estimate varies considerably across differentially and non-differentially expressed genes, we chose to address question **QB**. In this respect, two correlation sensitive statistics namely: Pearson's correlation (**P**) and Spearman rank correlation (**S**) [12] was used to segment pixels belonging to the foreground (**F**) and background (**B**), by statistically comparing adjacent rows and columns at a given significance level ($\alpha = 0.05$). The median value of the pixels in **F** and **B** was chosen as representative of the foreground and background intensities for that spot. The median is robust to outliers also termed as salt and pepper noise, hence its choice. As noted earlier, it is possible to encounter spots (Spot C) which have **F** = $\phi$ across both the channels. Such spots can be control spots with probes or poorly hybridized spots, hence uninteresting.

Correlation statistics provides a way to exclude such spots, hence useful in assessing spot quality prior to inferring differential expression. Spots which have $\mathbf{F} = \phi$ in only one of the channels may represent a gene that is expressed in only one of the channels, hence differentially expressed. For such spots, the median of all the pixels inside the grid in that channel ($\mathbf{F} = \phi$) was chosen as the foreground intensity. The algorithm for determining the $\mathbf{F}$ and $\mathbf{B}$ for a spot inside a grid is enclosed below.

*Determining Foreground and Background using Correlation Statistic*

*Algorithm:*

> **Given**: Spot inside a rectangular grid $\mathbf{I}$
>
> **Objective**: Generate a binary partition of $\mathbf{I}$ into foreground ($\mathbf{F}$) and background
>
> ($\mathbf{B}$) using correlation statistics.

**Step 1**: Given spot $\mathbf{I}$ consist of $m$ rows and $n$ columns (i.e. $m$ x $n$ pixels), represented by row vectors $R_i, i = 1...m$ and column vectors $C_j, j = 1...n$.

**Step 2**: Choose a measure of correlation $\Phi$ (such as Pearson's correlation or Spearman's correlation) and a significance value $\alpha = 0.05$. The p-value in the case of Pearson's correlation is determined by transforming the correlation into students t-statistics with n-2 degrees of freedom [12] where n represents the row/column size. A similar approach is used to determine the p-value of Spearman's correlation [12].

**Step 3**: Several corrections have been proposed in statistical literature. In the present study, each pair-wise comparison $\Phi(R_i, R_{i+1}), i = 1...m - 1$ is carried out independently of each other. Therefore, we use Bonferroni correction to control for family-wise error rate. The p-value obtained for the (*m-1*) pair-wise comparisons is compared to the adjusted significance level given by $\alpha^* = \alpha/(m\text{-}1)$. If the p-value is lesser than $\alpha^*$, the indices of the corresponding pairs are stored in set **R**.

**Step 4**: Repeat Step 2, with the column vectors, i.e. determine significantly pair-wise correlation between the columns $\Phi(C_j, C_{j+1}), j = 1...n-1$. Store the indices of the columns that were significantly correlated in set **C**.

**Step 5**: The foreground pixels are those that lie in the intersection set, given by **F** = **C** $\bigcap$ **R**. The background pixels are those which are in **I** but not in **F**, i.e. **B** = **I**\**F**.

**Step 6**: The foreground intensity is represented by the median of the pixels in **F** and the background intensity by median of the pixels in **B**

**Step 7**: Repeat the Steps 1, 2, 3, 4 and 5 for the channels Cy3 and Cy5 independently.

It is important to note that the above segmentation does not pose any constraint on the connectivity of the pixels comprising a spot. It also approximates seemingly circular spots by a rectangle.

A change in the expression of genes between the control and the experimental channels can be an outcome of either true biological variability or experimental artifacts. Several non-biological factors can contribute to differential expression. Normalization is the procedure of *minimizing* the effect of experimental artifacts and forms an important step prior to inferring differential gene expression. In the subsequent discussion the term normalization implies LOWESS normalized data [14, 15]. The raw data obtained after segmentation for each of the five replicate arrays was normalized using LOWESS regression prior to inferring differential expression.

## 3. Results

The foreground and the background intensities across the Cy3 and Cy5 channels across five replicate arrays in HDL metabolism experiment containing 19 x 21 = 399 spots [1, 12, 14] were estimated using five segmentation techniques, namely: k-means, PAM, SPOT, Pearson's correlation (**P**) and Spearman rank correlation (**S**). The performance of the segmentation techniques were subsequently assessed, this addresses question **QC**.

Visual inspection of the arrays Figure 1, indicate that majority of the spots on the array do not change significantly between the Cy3 and Cy5 channels. Spots that had $\mathbf{F} = \phi$ across the control and experimental channels on segmentation using the correlation statistics **P** and **S** were flagged as being poorly hybridized. Such spots cannot be quantitated, hence excluded from subsequent analysis. As expected, the number of flagged spots varied across replicate arrays, Figure 7. However, the profile of the flagged spots obtained using **P** and **S** did not change appreciably, Figure 7. Spots that were flagged even in one of the arrays were excluded from subsequent analysis as they were not reproducible. The number of spots which were not flagged across the five replicate arrays using correlation statistics **P** and **S** were 163 (~41%) and 149 (~ 37%). The median pixel intensities of the foreground region (**F**) corresponding to these spots were subsequently LOWESS normalized and used to infer differential gene expression. This has to be contrasted to SPOT, k-mean and PAM, where the median pixel intensities of the foreground region (F) corresponding to the 399 spots were LOWESS normalized and subsequently used to infer differential gene expression. Therefore, unlike k-means, PAM and SPOT, correlation statistics provides a way flag the spots prior to inferring differential gene expression, hence minimize false-positives. As expected, spot C was flagged but not spots A and B. The number of flagged spots is considerably high across **P** and **S**, however, this should not be surprising as only one out of the 399 spots (i.e. Apolipoprotein AI) was verified to be differentially expressed between the Cy3 and Cy5 channels [7, 13, 14].

In a recent study [7], t-statistics was proposed to determine differential gene expression and the performance of the various segmentation techniques. In the present study, we used parametric ttest to determine statistically significant differential expression ($\alpha$ = 0.05) on the LOWESS normalized data across the five replicate arrays obtained using the five segmentation techniques. The number of false-positives across each of the segmentation technique is shown in Figure 8. The number of false-positives picked up by the correlation statistics (**P** and **S**) was considerably lower than those picked by k-means, PAM and SPOT, Figure 8. This can attributed to the inherent feature of the correlation statistics that is useful in flagging the poorly hybridized spots. However, the gene Apolipoprotein A1 was identified as being differentially expressed by the five segmentation techniques and conforms to earlier studies [7, 10, 13, 14].

## 4. Discussion

Image segmentation forms a crucial preliminary step in microarray analysis as any errors incurred at this step is bound to propagate through subsequent analysis. Several image segmentation techniques were proposed in the past. To our knowledge the present study is the first of its kind where correlation of the pixels comprising a spot was investigated. The nature of correlation of microarray spots was investigated using Morgera's covariance complexity. Subsequently, correlation statistics namely: Pearson and Spearman correlation were used to segment the microarray spots. Correlation statistics was also shown to be useful in flagging poorly hybridized spots hence minimizing false-positives unlike SPOT, k-means and PAM. The present study also raises the intriguing question whether a change in correlation between the two channels can be an indicator of differential expression. Alternately, pixel correlation may be directly proportional to dye-binding. The results obtained using correlation statistics were compared to those obtained using k-means, PAM and SPOT segmentation techniques. Normalization is an integral part of microarray analysis and was included for completeness. The effectiveness of the segmentation techniques was tested on publicly available microarrays. The correlation statistics was found to be useful in minimizing the number of false-positives compared to other segmentation techniques as reflected by the t-statistics.

**Acknowledgments**:

We would like to thank Terry Speed's group at University of California at Berkeley for making available the microarray images and data generated in the Lipoprotein experiment. We would also like to thank the reviewers for their comments and suggestions.

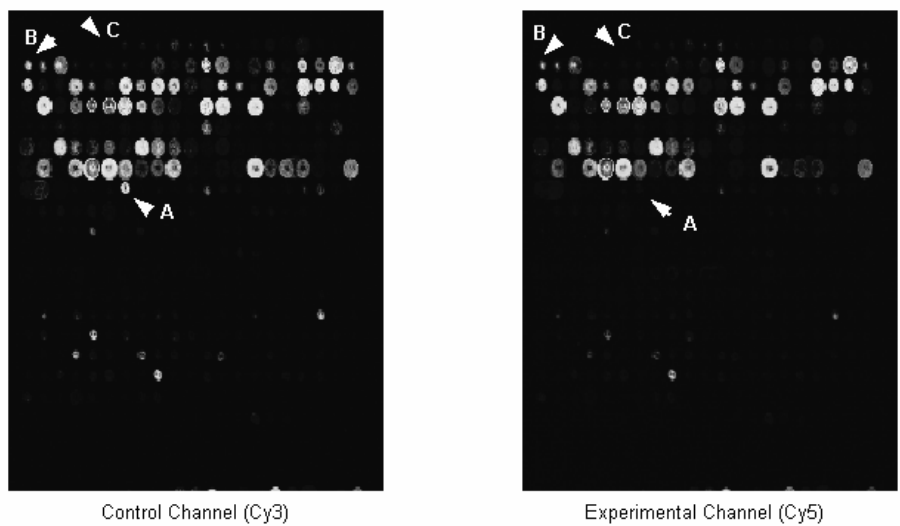

Control Channel (Cy3)                    Experimental Channel (Cy5)

**Figure 1**: Microarray scanned at two different wavelengths corresponding to Cy3 (~ 530 nm, control channel) and Cy5 (~ 630 nm, experimental channel). Spots A, B and C are indicated by solid arrows. Spot A represents the gene Apo AI which is down-regulated. Spot B represents an EST that is expressed equally in both the channels. Spot C represents a BLANK spot that is neither expressed in the control nor the experimental channel.

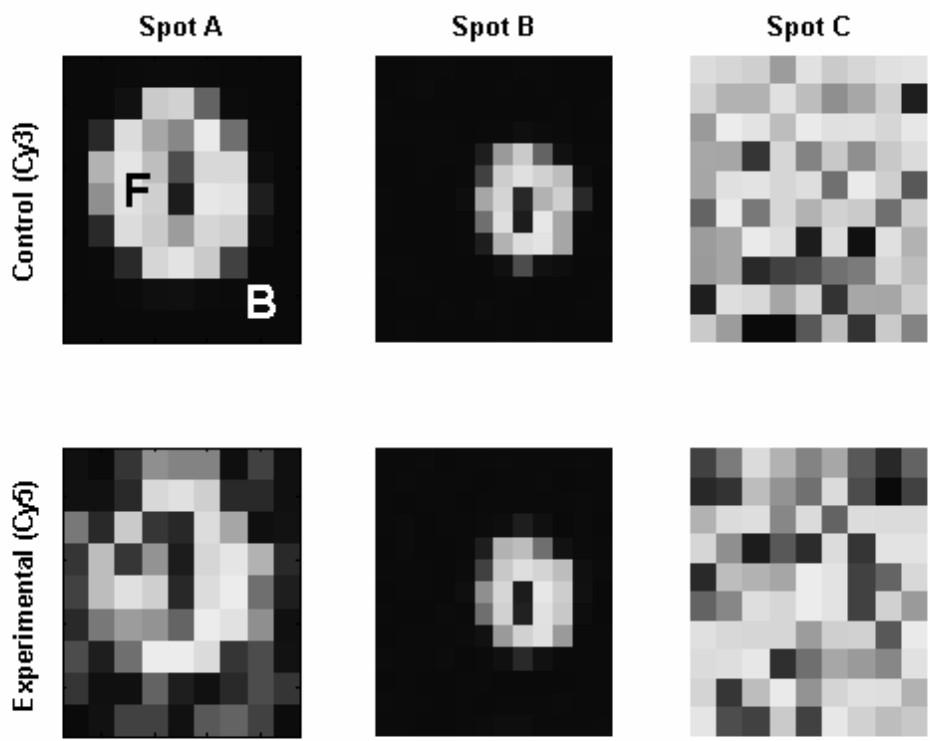

**Figure 2** Pixel intensities of spots located in a grid for one of the arrays. Spot A represents a gene that is differentially expressed between the control (Cy3, top) and the experimental (Cy5, bottom) channels. Spot B represents a gene that is equally expressed in the control and the experimental channels. Spot C represents a gene that is neither expressed in the control nor the experimental channel. **F** and **B** in (a) correspond to the foreground and the background regions.

*Popular Microarray Image Segmentation Algorithms*

Key: Foreground: (**F**); Background: (**B**); Target Intensity: Estimate of **F** – Estimate of **B**

**A. Fixed Circle**

> **F**: Pixels inside the fixed circle
> **B**: Pixels not inside the fixed circle but inside the grid
> *Assumptions*: Foreground region is circular in shape and is constant across all spots.
> Diameter is user specified.
> *Implemented by*: ScanAlyze: rana.lbl.gov
> GenePix: www.axon.com
> Scanarray: las.perkinelmer.com
> *Reference: 8*

**B. Adaptive Circle**

> **F**: Pixels inside a circle
> **B**: Pixels from the *valley spot* which are consists of representative pixels from the furthest four corners of the given spot.
> *Assumptions*: Foreground region is circular in shape but not constant across all spots.
> Existence of valley spot. Diameter user-specified.
> *Implemented by*: GenePix: www.axon.com
> *Reference: 7*

**C. Adaptive Shape**

> **F**: Pixels inside the region determined by the region growing approach
> **B**: Determined by morphological opening.
> *Assumptions*: Seed is user-specified.
> *Implemented by*: Spot: www.cmis.csiro.au
> *Reference: 7*

**D. Histogram**

> **F**: Pixels greater than threshold determined by non-parametric Mann-Whitney test
> **B**: All other pixels outside the target mask
> *Assumptions*: pre-defined target mask, statistical testing on ranks as opposed to the true values.
> *Implemented by*: Scanarray: las.perkinelmer.com
> *Reference: 9, 10*

**Figure 3** Popular microarray image segmentation algorithms.

*Morgera's Covariance Complexity*

**Given**: Spot inside a rectangular grid (**I**)

**Objective**: Determine the extent of correlation of the pixels comprising the spot.

**Step 1**: Determine the eigen-values, by singular value decomposition (SVD) [16] of the matrix **I**, $I_{ij}, i = 1...m, j = 1...n$, to yield eigen-values $l_i, i = 1...p$, where $p = \min(m, n)$.

The eigen-values $l_i^2, i = 1...p$ can also be determined by eigen-decomposition of the symmetric matrices $\mathbf{I}^T\mathbf{I}$ and $\mathbf{II}^T$. While the former ($\mathbf{I}^T\mathbf{I}$) captures the column-wise correlation, the latter ($\mathbf{II}^T$) captures the row-wise correlation.

**Step 2**: The normalized variance along the $i^{th}$ component is given by

$$s_i = \frac{(l_i)^2}{\sum_{i=1}^{p}(l_i)^2}, i = 1...p$$

**Step 3**: Morgera's covariance complexity (**$h$**) [11] is given by

$$h = -\frac{1}{\log p} \sum_{k=1}^{p} s_k \log s_k$$

The value of (**$h$**) lies in the interval $0 \le h \le 1$ and is *inversely* proportional to the pixel correlation.

**Figure 4** Morgera's covariance complexity

Given Image **I**

[Image: grid with values 1200, 800, 400, 500 / 900, 700, 600, 300 / 1100, 1300, 1000, 850]

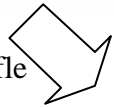

Row-wise Shuffle

[Image: grid with values 900, 700, 600, 300 / 1200, 800, 400, 500 / 1100, 1300, 1000, 850]

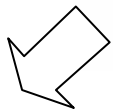

Column-wise shuffle

Constrained randomized shuffle (**I\***)

[Image: grid with values 700, 900, 300, 600 / 800, 1200, 500, 400 / 1300, 1100, 850, 1000]

**Figure 5** Generating a constrained randomized shuffled image (**I\***) from the given image (**I**).

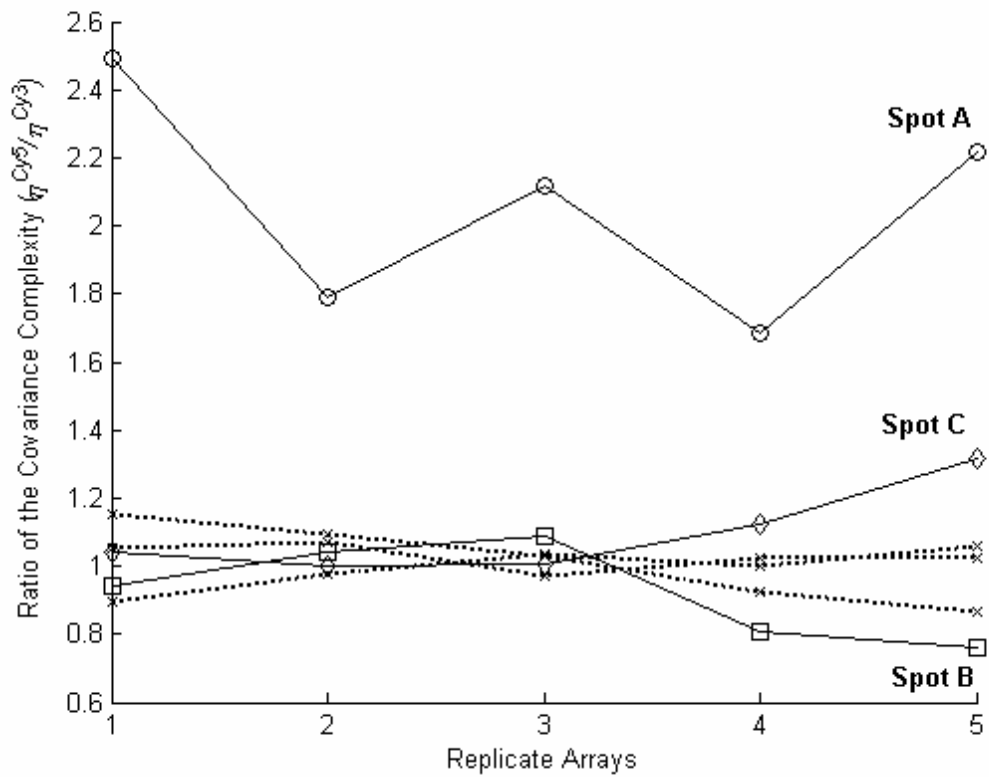

**Figure 6** Ratio of the covariance complexity between the Cy3 and Cy5 channels ($\boldsymbol{h}^{\text{Cy5}}/\boldsymbol{h}^{\text{Cy3}}$) for Spot A which is differentially expressed (circle), for Spot B which is not differentially expressed but exhibits significant correlation across the two channels (diamond) and for Spot C which is not differentially expressed and does not exhibit significant correlation across both the channels (square) across the five replicate arrays. The ratio of the covariance complexity on the corresponding random shuffled counterparts is represented by dotted lines for the three spots across five replicate arrays (x). The random shuffled counterparts correspond to spots which fail to show sufficient correlation across both the channels.

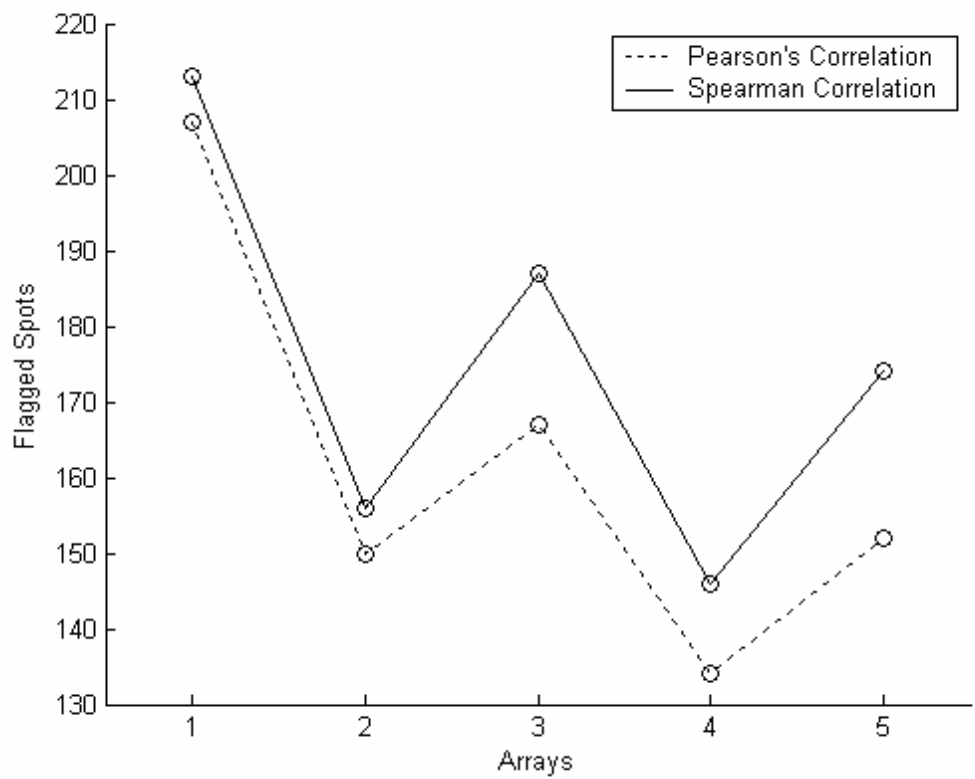

**Figure 7** The number of flagged spots obtained using Pearson's and Spearman rank correlation exhibit similar profile across the five replicate arrays.

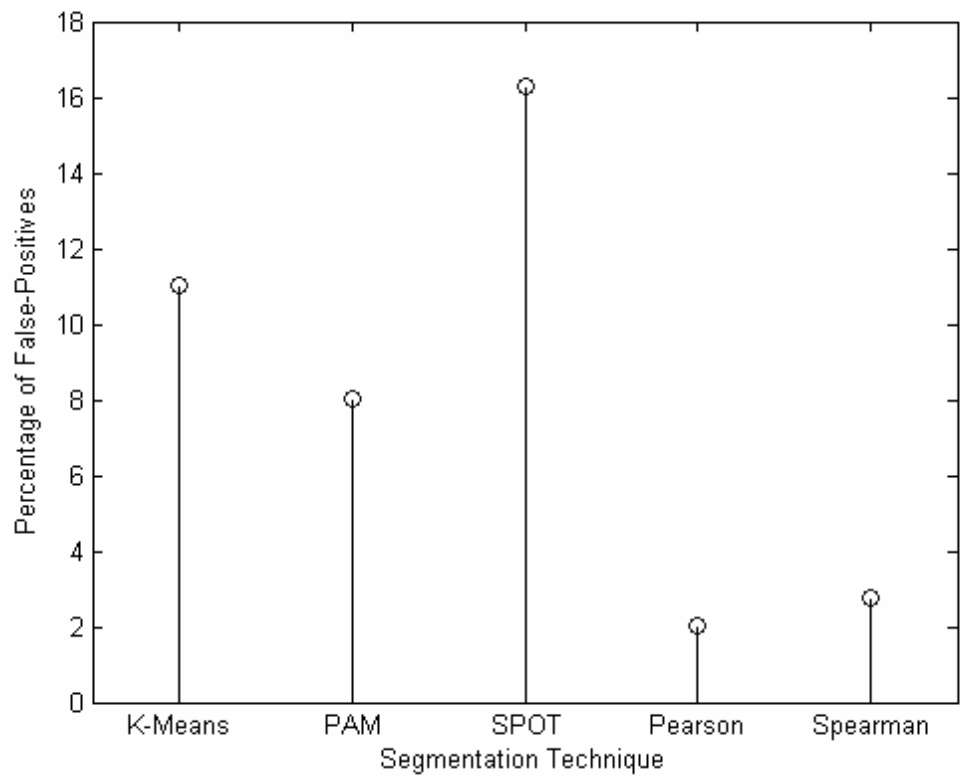

**Figure 8** Number of false positives across each of the segmentation techniques.